\documentclass[%
reprint,
unsortedaddress,
showpacs,preprintnumbers,
amsmath,amssymb,
pra,
longbibliography,
lengthcheck,%
]{revtex4-1}

\usepackage{graphicx}
\usepackage{dcolumn}
\usepackage{bm}
\usepackage{subfigure}

\begin{document}

\preprint{APS/123-QED}

\title{Entanglement in (1/2,1) mixed-spin XY model with long-range interactions}

\author{Seyit Deniz Han$^1$}
\author{Tu\^{g}ba T\"{u}fek\c{c}çi$^1$}
\author{Timothy P. Spiller$^2$}
\author{Ekrem Aydiner$^1$}
\email{ekrem.aydiner@istanbul.edu.tr} \affiliation{$^1$Istanbul
University Theoretical Physics Research Group, \.{I}stanbul University, Tr-34134, \.{I}stanbul, Turkey \\
$^2$Quantum Information Science, School of Physics and Astronomy,
University of Leeds, Leeds, LS2 9JT, United Kingdom}


\date{\today}

\begin{abstract}
In this study, considering the long-range interaction with an
inverse-square and its trigonometric and hyperbolic variants in SCM
model we investigate entanglement in (1/2,1) mixed-spin XY model. We
also discuss the temperature and magnetic field dependence of the
thermal entanglement in this system for different types of
interaction. The numerical results show that, in the presence of the
long-range interactions, thermal entanglement between spins has a
rich behavior dependent upon the interaction strength, temperature
and magnetic field. Indeed we find that for less than a critical
distance there are entanglement plateaus dependent upon the distance
between the spins, whereas above the critical distance the
entanglement can exhibit sudden death.
\end{abstract}


\maketitle


\section{\label{sec:intro}Introduction}

Over the last decade or so, quantum entanglement has been recognized
as crucial in various fields of quantum information \cite{Nielsen}
such as in quantum computation \cite{Schumacher}, quantum
teleportation \cite{Bennett1}, superdense coding \cite{Bennett2},
quantum communication \cite{Divincenzo}, quantum perfect state
transfer \cite{Christandl}, quantum cryptology \cite{Ekert,Deutsch}
and quantum computational speed-ups \cite{Shor,Grover}. Potential
applications of entanglement in these fields have stimulated
research on methods to quantify and control it. In the solid state
arena, considerable attention has been devoted to interacting
Heisenberg spin systems. Spins are recognized as candidates for
solid state quantum computation
\cite{Loss,Burkard,Imamoglu,Platzman, Raussendorf} and even short
range communication \cite{Bose,Kay}. Interactions such as Heisenberg
enable gate operations in solid state quantum computation processors
\cite{Loss,Burkard,Imamoglu,Platzman, Raussendorf}. Therefore
significant research has been performed to understand quantum
entanglement behavior in spin systems, such as all various kinds of
Heisenberg (XX, XY, XXZ and XYZ) models, and similarly Ising models.
Many detailed and extensive investigations have addressed the
dependence of entanglement on parameters such as magnetic field,
single-ion anisotropy, temperature and interaction strength
\cite{Kamta1,Yeo,Kamta2,SunY,Zhang,WangX,Arnesen,Zhang1,Asoudeh,
Santos,Zhang2,Gu,Albayrak1,Ercolessi,Liu,Zhou,Abliz,Chen,Li,SunZ,Zhang3,Zhang4,Akyuz1,WangX2,WangX3}.

Recently, a new type of long-range interaction has been used in
Refs.\,\cite{LiB,Gaudiano,Giuliano} to obtain long-distance
entanglement in the spin system. In these works, spin pair
interaction is given with a factor inversely proportional to a power
of the distance between sites as $J\left( R\right) \sim R^{-\alpha
}$. It is shown in these studies that long-distance entanglement can
be obtained using this interaction type for different values of the
interaction parameter $\alpha$ in the Heisenberg spin systems. Here
we must remark that this interaction type and its variants except
the entanglement have been considered to explore many physical
phenomena. Indeed, the inverse-square, trigonometric and hyperbolic
interacting particle systems \cite{Calogero,Sutherland,Moser} and
its spin generalizations
\cite{Gibbons,Wojciechowski,Haldane1,Shastry1} are important model
of many-body systems due to their exactly solvability and intimate
connection to spin systems in condensed matter
\cite{Gibbons,Wojciechowski,Haldane1,Shastry1,Haldane2,Shastry2,
Kawakami,Hikami,Fowler,Basu,Polychronakos}, and in the other areas
in physics. These interaction types are called different names such
as Haldane-Shastry models, Calogero-Moser or
Sutherland-Calogero-Moser models or interactions in the literature.
But, it is assumed that generalized Haldane-Shastry models is the
supersymmetric partners of the Sutherland-Calogero-Moser type
models. Hence, for historical reasons, in this study we will express
these interactions as Sutherland-Calogero-Moser (SCM) model or SCM
tpye interactions.

In this study, inspired by Refs.\,\cite{LiB,Gaudiano,Giuliano}, we
will consider long-range interaction with an inverse-square and its
trigonometric and hyperbolic variants given in the SCM model to
obtain entanglement between two spin in the mixed-spin (1/2,1) XY
Heisenberg spin system. Our numerical results show that the
long-range interactions lead to topological plateaus in
entanglement. That is, the entanglement between spins still subsists
throughout plateau up to critical distance although spins are
separated from each other.

This paper is organized as follows. In Section II we define the
mixed spin-1/2 and spin-1 XY model, and give the eigenvalues,
eigenvectors, density matrix of the system and definition of
negativity as a measure of entanglement. In Section III, we present
and discuss the numerical results of modeling, giving the negativity
of the system for SCM types i.e. inverse-square and its
trigonometric and hyperbolic variants with varying model parameters.
Finally Section IV is devoted to conclusions.

\section{\label{sec:model}Model}
The Hamiltonian for the isotropic mixed spin (1/2,1) Heisenberg XY
spin chain under an applied magnetic field $B$ is given by
\begin{equation}
H=\sum_{i}
B(s_{i}^{z}+S_{i+1}^{z})-J(R)\sum_{i}[s_{i}^{x}S_{i+1}^{x}+s_{i}^{y}S_{i+1}^{y}]
\end{equation}
where $J(R)$ is the spin interaction coupling which will be defined
in terms of SCM interactions. Also $s_{i}$ and $S_{i}$ ($i=1,2$) are
spin operators of the spin-1/2 and spin-1 components, respectively.
The basis vectors of the two-spin system defined in Eq.\,(1) are
$\{$ $\left\vert -1/2,-1\right\rangle$, $\left\vert
1/2,0\right\rangle$, $\left\vert -1/2,1\right\rangle$, $\left\vert
1/2,-1\right\rangle$, $\left\vert -1/2,0\right\rangle$, $\left\vert
1/2,1\right\rangle$ $\}$, where $\left\vert s,S\right\rangle $ is
the eigenstate of $s^{z}$ and $S^{z}$ with corresponding eigenvalues
given by $s$ and $S$, respectively. The corresponding eigenvalues
and eigenvectors of $H$ for all $J(R)$ can be given as follows
\begin{subequations}
\begin{eqnarray}
H\left\vert -1/2,-1\right\rangle =-3/2 B\left\vert
-1/2,-1\right\rangle
\\
H\left\vert 1/2,1\right\rangle =3/2B\left\vert 1/2,1\right\rangle
\\
H\left\vert \phi ^{\pm }\right\rangle =W_{\pm }\left\vert \phi ^{\pm
}\right\rangle
\\
H\left\vert \psi ^{\pm }\right\rangle =Q_{\pm }\left\vert \psi ^{\pm
}\right\rangle
\end{eqnarray}
\end{subequations}
where
\begin{subequations}
\begin{eqnarray}
\phi ^{\pm }=1/ \sqrt{2}(\left\vert 1/2,0\right\rangle \pm
\left\vert 1/2,-1\right\rangle )
\\
\psi ^{\pm }=1/ \sqrt{2}(\left\vert -1/2,1\right\rangle \pm
\left\vert -1/2,0\right\rangle ) .
\end{eqnarray}
\end{subequations}
Here $\phi ^{\pm }$ and $\psi ^{\pm }$ are maximally entangled
states---so-called Bell states. The eigenstates of the isotropic XY
model given by Eq.\,(2) with Eq.\,(3) do not change for different
$J(R)$ values. However, the eigenvalues $W_{\pm }$ and $Q_{\pm }$ in
Eq.\,(2) do change dependent upon the value of $J(R)$. Each
eigenvalue, for different types of the SCM interaction, will be
explored in detail in the next section.

The partial transpose of the density matrix $\rho$ for the two-spin
which is defined by Eq.\,(1) is written in the form
\begin{equation}
\rho ^{T_{1}}=\left(
\begin{array}{cccccc}
a_{11} & a_{12} & 0 & 0 & 0 & 0 \\
a_{21} & a_{22} & 0 & 0 & 0 & 0 \\
0 & 0 & a_{33} & 0 & 0 & 0 \\
0 & 0 & 0 & a_{44} & 0 & 0 \\
0 & 0 & 0 & 0 & a_{55} & a_{56} \\
0 & 0 & 0 & 0 & a_{65} & a_{66}%
\end{array}%
\right)
\end{equation}
where $a_{ij}$ ($i,j=1,...,6)$ are matrix elements of $\rho
^{T_{1}}$.  For different cases of $J$, the matrix elements of $\rho
^{T_{1}}$ are presented, along with the negativity.

Negativity is a measure of the quantum entanglement appropriate for
application to higher spins \cite{Peres,Horodecki,Vidal}. Therefore
we will use the concept of negativity to study entanglement in
(1/2,1) mixed-spin XY systems. The negativity can be obtained using
the density operator of the quantum system. The state of a system at
thermal equilibrium can be described by the density operator $\rho
\left( T\right) =\exp \left( -\beta H\right) /Z$, where $Z=Tr\left(
\exp \left( -\beta H\right) \right) $ is the partition function and
$\beta =1/kT$ ($k$ is Boltzmann's constant, which is set to unity
$k=1$ hereafter for the sake of simplicity, and $T$ is the
temperature). By choosing an appropriate set of orthonormal product
basis states $\left\{ \left\vert v_{i}v_{j}\right\rangle \right\}
\equiv \left\{ \left\vert v_{i}\right\rangle \left\vert
v_{j}\right\rangle \right\} $ for the density operator, the partial
transpose is defined by its matrix elements
\begin{equation}
\rho _{m\mu ,n\nu }^{T_{1}}=\left\langle v_{m}v_{\mu }\right\vert
\rho \left\vert v_{n}v_{\nu }\right\rangle =\rho _{m\nu ,n\mu } .
\end{equation}
The negativity of a state $\rho$ is by definition
\begin{equation}
N=\sum_{i}\left\vert \mu _{i}\right\vert
\end{equation}
where $\mu _{i}$ is the negative eigenvalue of the partial transpose
density matrix $\rho ^{T_{1}}$. Here $\rho ^{T_{1}}$ is the partial
transpose with respect to the first system. The negativity $N$ is
related to trace norm of $\rho ^{T_{1}}$ via
\begin{equation}
N=\frac{\left\Vert \rho ^{T_{1}}\right\Vert -1}{2}
\end{equation}
where the trace norm of $\rho ^{T_{1}}$ is equal to the sum of the
absolute values of the eigenvalues of $\rho ^{T_{1}}$.

\section{\label{sec:results}Results and Discussions}

In this section in order to discuss entanglement we will compute the
negativity to quantify entanglement for two-spin in the isotropic
mixed-spin (1/2,1) XY Heisenberg spin chain given by Eq.\,(1)
considering different long-range interaction types given in the SCM
model
\cite{Calogero,Sutherland,Moser,Gibbons,Wojciechowski,Haldane1,Shastry1}.
Therefore, in the below Subsections, we will consider inverse-square
interaction coupling $J_{0}/R^2$ as type I, trigonometric
interaction coupling $J_{0}/\sin ^{2}(R)$ as type II and hyperbolic
interaction coupling $J_{0}/\sinh ^{2}(R)$ as type III in
Subsections III.A,B and C respectively.

\subsection{SCM Type I: $J(R)=J_{0}/R^2$ }

Firstly we consider the inverse-square interaction type in SCM
model, which is defined with exchange interaction $J(R)=J_{0}/R^2$
\cite{Calogero,Sutherland,Moser,Gibbons,Wojciechowski,Haldane1,Shastry1}.
Here we set $J_{0}=1$ for simplicity. The eigenvalues for this type
of SCM interaction are given by
\begin{eqnarray}
\pm 3B/2, \quad W_{\pm }=-\frac{B}{2}\pm \frac{\sqrt{2}}{2R^2},
\quad Q_{\pm }=\frac{B}{2}\pm \frac{\sqrt{2}}{2R^2}.
\end{eqnarray}
Using the eigenvalues in Eq.\,(8) and the eigenvectors in Eq.\,(2)
(with Eq.\,(3)) the density matrix $\rho$ can be constructed.  For a
SCM interaction of type $J(R)=1/R^2$, the partial transpose of
$\rho$ shown in Eq.\,(4) has matrix elements given in Eq.\,(9) as
follows:
\begin{subequations}
\begin{eqnarray}
a_{11}=\frac{1}{Z}e^{\frac{3B}{2T}}
\\
a_{12}=a_{21}=-\frac{1}{Z}e^{\frac{B}{2T}}\sinh
[\frac{1}{\sqrt{2}R^2T}]
\\
a_{22}=a_{33}=\frac{1}{Z}e^{-\frac{B}{2T}}\cosh
[\frac{1}{\sqrt{2}R^2T}]
\\
a_{44}=a_{55}=\frac{1}{Z}e^{\frac{B}{2T}}\cosh
[\frac{1}{\sqrt{2}R^2T}]
\\
a_{56}=a_{65}=-\frac{1}{Z}e^{-\frac{B}{2T}}\sinh
[\frac{1}{\sqrt{2}R^2T}]
\\ a_{66}=\frac{1}{Z}e^{-\frac{3B}{2T}}
\end{eqnarray}
\end{subequations}
where the partition function $Z$ is written as
\begin{equation}
Z=2\left\{ \cosh [\frac{3B}{2T}]+2\cosh [\frac{B}{2T}]\cosh [\frac{1}{\sqrt{%
2}R^2T}]\right\} \ .
\end{equation}
For the SCM interaction of type $J(R)=1/R^2$, the negativity given
by Eq.\,(7) for the two-spin system defined in the Heisenberg mixed
XY model in Eq.\,(1) can obtained from summation of the negative
eigenvalues of the partial transpose matrix Eq.\,(4) with elements
as in Eq.\,(9).

The negativity of the two-spin system in the case $J(R)=1/R^2$ is
plotted versus $R$ for different temperature $T$ values at fixed
magnetic field $B=1$ and for different magnetic field values at
fixed temperature $T=1$, in Fig.\,(1a) and Fig.\,(1b), respectively.
Firstly we note that there is a singularity at $R=0$, hence, as it
can be seen from both figures that negativity drops to zero in the
limit $R\rightarrow 0$. On the other hand, as it can be seen from
Fig.\,(1a), the negativity reaches a plateau for small $R$ at a
fixed temperature $T$  and fixed magnetic field $B$. The value of
the negativity plateau is 0.5 at very low temperatures, decreasing
at higher temperatures although with the plateau behavior still
apparent. Similar behavior is observed in Fig.\,(1b), with the
negativity attaining a plateau at small $R$. For relatively high
magnetic fields the plateau value approaches 0.5, but decreases for
lower magnetic values as expected. A very interesting aspect of the
behavior in these figures is that the negativity shows this plateau
behavior, like in topological quantization, dependent upon the
distance between the two spins for different temperatures at fixed
$B$ and similarly for different magnetic fields at fixed
temperature. It is interesting to note that the plateau width (in
$R$) increases as it approaches its maximum for lowering $T$,
whereas the width decreases as it approaches its maximum for
increasing $B$. In both figures, for increasing $R$, in some cases
the negativity exhibits a form of sudden death \cite{yu1,yu2,yu3},
reaching zero at a critical $R$ value of $R_c$ \cite{sdcomment}. The
results of Fig.\,(1a) demonstrate the dependence of the critical
$R_c$ on temperature $T$ at fixed $B$, whereas the results of
Fig.\,(1b) demonstrate some insensitivity of $R_c$ to magnetic field
$B$ at fixed $T$ but with the entanglement slope (at $R_c$, as a
function of $R$) dependent upon $B$. Here we express that the
topological plateaus in Fig.\,(1) emerge from the presence of the
long-range interaction with inverse-square $1/R^2$ in this model.
\begin{figure}
\centering{
\includegraphics[width=3.5in,height=2.5in]{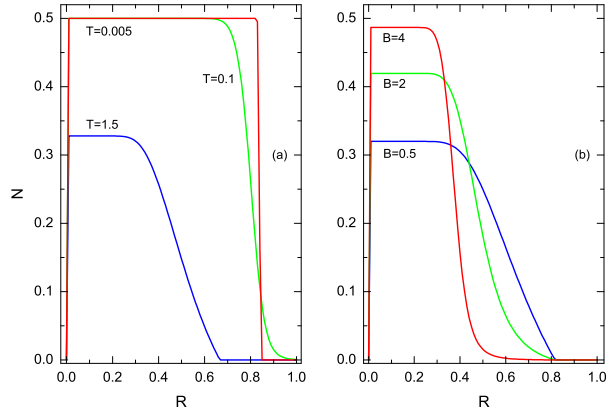}
\caption{Negativity as a function of $R$ for the two-spin
ferromagnetic XY system (a) for selected temperatures at fixed
magnetic field $B=1$, (b) for selected external uniform magnetic
fields at fixed temperature $T=1$.}}
\end{figure}
\begin{figure}
\centering{
\includegraphics[width=3.5in,height=2.5in]{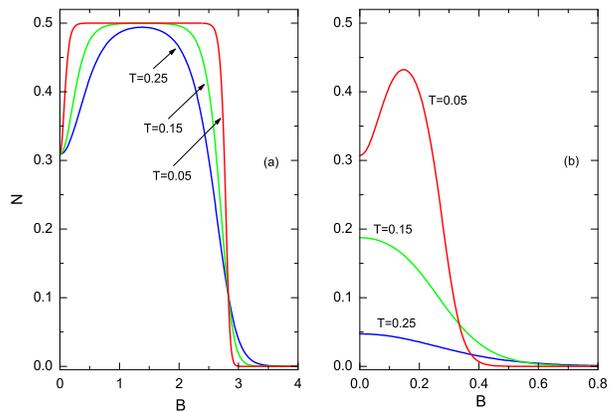}
\caption{Magnetic field dependence of negativity in the two-spin
ferromagnetic XY system for selected temperatures (a) at fixed
$R=0.5$, (b) at fixed $R=1.5$.}}
\end{figure}

In order to further investigate the magnetic field dependence of the entanglement for
different $R$ values in the larger $R$ regime, the negativity is plotted as a function of
magnetic field for different temperature values ($T=0.05, 0.15,
0.25$), at fixed $R=0.5$ in Fig.\,(2a),  and at fixed
$R=1.5$ in Fig\,(2b). It can be seen from these figures that although the negativity
curves have different detailed characteristics, the negativity for the two
different values of $R$ tends to zero for increasing $B$ values. Identifying a critical $B_c$ beyond which there is no entanglement suggests that this critical $B_c$ value and the manner in which the entanglement vanishes both depend upon the chosen
$R$ and the temperature $T$.

Similarly, to further investigate the temperature dependence of the
entanglement for different $R$ values in the larger $R$ regime, the
negativity is plotted versus temperature for different magnetic
field values ($B=0.5, 2, 4$), at fixed $R=0.5$ in Fig.\,(3a), and
for different magnetic field values ($B=0.05, 0.15, 0.35$) at fixed
$R=1.5$ in Fig.\,(3b). Since the negativity does not appear for
$B=0.5, 2, 4$ at $R=1.5$, we plot Fig.\,(3b) for different $B$
values. As it can be seen from these figures, the temperature
dependence of the negativity for different $R$ and $B$ values has
very similar characteristic behavior.  In both figures, the
negativity curves for different magnetic values meet at the same
temperature point which indicates a critical temperature value
$T_c$. Furthermore, Fig.\,(3) demonstrates critical temperature
$T_c$ dependence on $R$, consistent with the critical $R_c$
dependence on $T$ of Fig.\,(1a). Similarly to Fig.\,(1) singularity
behavior in negativity appears in the limit $R\rightarrow 0$ in
Fig.\,(3).
\begin{figure}
\centering{
\includegraphics[width=3.5in,height=2.5in]{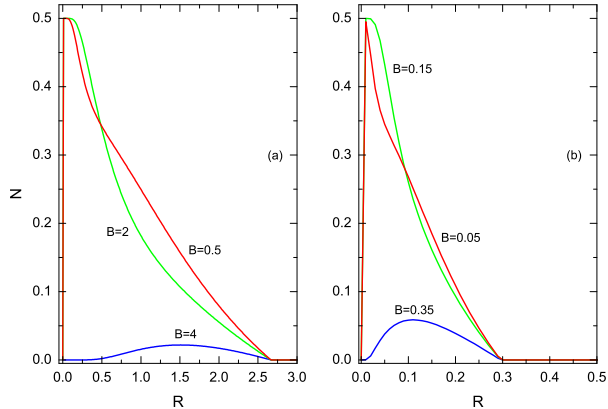}
\caption{Temperature dependence of the negativity in the two-spin ferromagnetic
XY system for selected magnetic field values (a) at fixed $R=0.5$, (b) at
fixed $R=1.5$.}}
\end{figure}
\begin{figure}
\centering{\includegraphics[width=3.0in,height=2.5in]{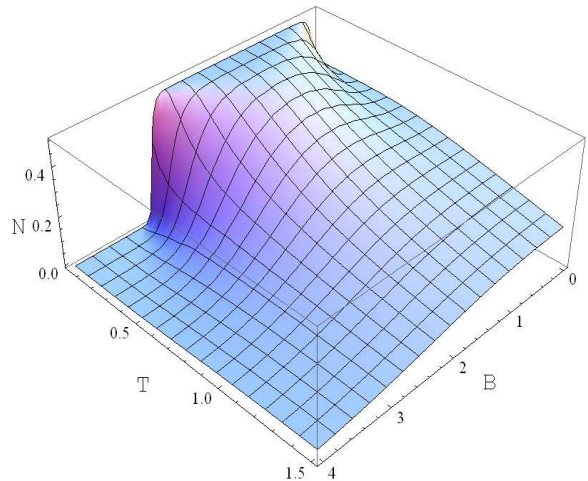}
\caption{Surface plot of the negativity in the two-spin
ferromagnetic XY system as a function of temperature and magnetic
field at fixed $R=0.5$}}
\end{figure}

To further demonstrate and unify the dependence of the entanglement
on temperature and magnetic field, a surface plot of the negativity
is given in Fig.\,(4), for an example fixed $R=0.5$, as a function
of both $T$ and $B$. As can be seen from Fig.\,(4), the negativity
of the two-spin XY system has a plateau for this fixed $R$ value. We
also know from the previous Figs.\,(1), (2) and (3) that the range
of the plateau depends on magnetic field and temperature and we can
qualitatively comment that range (in $T$ and $B$) of the plateau for
small $R$ values will be bigger than that for large $R$ values.
Furthermore, Fig.\,(4) shows clearly that for very low temperature
values the negativity exhibits critical behavior (coming off the
plateau) and decreases rapidly with increasing field at a critical
magnetic field value. However, for higher temperatures beyond $T_c$
when there is no access to the plateau, the negativity simply
decreases smoothly for increasing magnetic field.

To summarize, in this subsection, we have investigated the behavior
of the entanglement of a two-spin XY system in the case of
long-range SCM interaction $J(R)=1/R^2$. We have discussed the
dependence of the negativity on interaction parameter $R$, magnetic
field $B$ and temperature $T$. We have shown that entanglement
appears in the presence of the long-range interaction in some
parameter regimes, and there is critical behavior in entanglement
which suddenly deaths at critical values of interaction parameter,
magnetic field and temperature.

\subsection{SCM Type II: $J(R)=J_{0}/\sin ^{2}R$ }

Secondly we consider trigonometric version of long-range interaction
in SCM model, which is defined with exchange interaction
$J(R)=J_{0}/\sin ^{2}R$
\cite{Calogero,Sutherland,Moser,Gibbons,Wojciechowski,Haldane1,Shastry1}.
Here we set $J_{0}=1$ and $\theta=R$ for simplicity. The eigenvalues
for this type of SCM interaction are given by
\begin{eqnarray}
\pm 3B/2, W_{\pm }=\frac{\mp B\varphi +4\sqrt{\varphi }}{2\varphi },
Q_{\pm }=\frac{\mp B\varphi -4\sqrt{\varphi }}{2\varphi } \quad
\end{eqnarray}
where $\varphi =3-4\cos [2R]+\cos [4R]$. Using the eigenvalues in
Eq.\,(11) and the eigenvectors in Eq.\,(2) (with Eq.\,(3)) the
density matrix $\rho$ can be constructed. For the SCM interaction
type of $J(R)=1/\sin ^{2}R$, the partial transpose of $\rho$ shown
in Eq.\,(4) has matrix elements given in Eq.\,(12) as follows:
\begin{subequations}
\begin{eqnarray}
a_{11}=\frac{1}{Z}e^{\frac{3B}{2T}}
\\
a_{12}=a_{21}=\frac{\mu_{-} }{2Z}(-e^{-\delta_{1}}+e^{-\delta_{2}})
\\
a_{22}=a_{33}=\frac{\mu_{+} }{2Z}(e^{\delta_{1}}+e^{\delta_{2}})
\\
a_{44}=a_{55}=\frac{\mu_{-}}{2Z}(e^{-\delta_{1}}+e^{-\delta_{2}})
\\
a_{56}=a_{65}=\frac{\mu_{+}}{2Z}(e^{\delta_{1}}-e^{\delta_{2}})
\\
a_{66}=\frac{1}{Z}e^{-\frac{3B}{2T}}
\end{eqnarray}
\end{subequations}
where $\delta_{1}=\frac{B\cos [2R]\csc [R]^{4}}{2T}$, $\delta_{2}=\frac{%
B(3+\cos [4R])\csc [R]^{4}}{8T}$ and $\mu_{\mp} =\exp(\mp\frac{\csc
[R]^{4}(3B+4B\cos [2R]+B\cos [4R]+8\sqrt{2}\sqrt{\sin
[R]^{4}})}{16T})$. The partition function $Z$ for the interaction type
  $J=1/\sin ^{2}(R)$ can therefore be written as
\begin{eqnarray}
Z=\mu _{-}e^{-\left( \delta _{1}+\delta _{2}\right) }\{e^{\delta
_{1}+2\delta _{2}}+e^{2\delta _{1}+\delta _{2}}+\mu
_{+}e^{\frac{5\delta _{1}-\delta _{2}}{2}}\\ \nonumber +\mu
_{+}^{2}e^{\delta _{2}}+\mu _{+}e^{\frac{-\delta _{1}+5\delta
_{2}}{2}}+\mu _{+}^{2}e^{\delta _{1}}\} \ .
\end{eqnarray}
For the SCM interaction of type $J(R)=1/\sin ^{2}R$, the negativity
given by Eq.\,(7) for the two-spin system defined in the Heisenberg
mixed XY model in Eq.\,(1) can obtained from summation of the
negative eigenvalues of the partial transpose matrix Eq.\,(4) with
elements as in Eq.\,(12).

The negativity for this $J(R)=1/\sin ^{2}R$ case is plotted versus
$R$ ($0\leq R\leq \pi $) for different temperature $T$ values at
fixed magnetic field value $B$ in Fig.\,(5a), and for different
magnetic field values at fixed temperature $T$ in Fig.\,(5b). One
can see that negativity drops to zero in the limit $R\rightarrow 0$
and $R\rightarrow \pi$ in both figures. In these limits, the
negativity have two singularities. On the other hand, it can be seen
from these figures that the negativity drops suddenly to zero with
$R$, creating a valley symmetric about $R=\pi/2$. This occurs for a
fixed $B=1$ value at the various temperature values used for
Fig.\,(5a), with the valley width and side plateau heights dependent
upon the chosen $T$ value and lower $T$ creating a narrower,
higher-sided ($N=0.5$) valley. The effect also occurs for a fixed
$T=1$ value at the various magnetic field values used for
Fig.\,(5b), with the valley width and side plateau heights dependent
upon the chosen $B$ value and higher $B$ creating a slightly wider,
steeper-sided and higher-sided (approaching $N=0.5$) valley. Given
the symmetric behavior of the negativity in the interval $0\leq
R\leq \pi$, as shown in Figs.\,(5) it is clear that there are two
critical $R_c$ values in these figures, equidistant from $R=\pi/2$.
\begin{figure}
\centering{
\includegraphics[width=3.5in,height=2.5in]{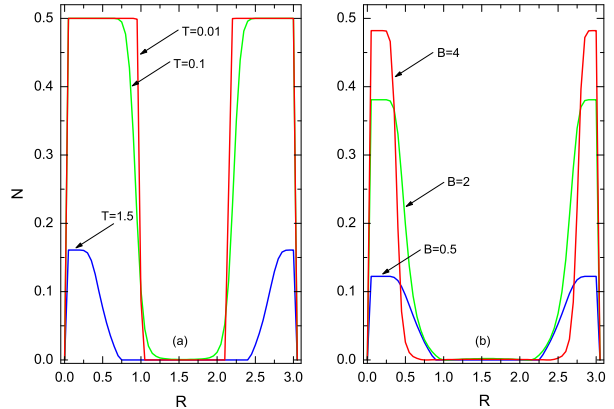}
\caption{Negativity as a function of $R$ for the two-spin
ferromagnetic XY system in the case $J(R)=1/\sin ^{2}R$ (a) for
selected temperatures at fixed magnetic field $B=1$, (b) for
selected external uniform magnetic fields at fixed temperature
$T=1$.}}
\end{figure}
\begin{figure}
\centering{
\includegraphics[width=3.0in,height=2.5in]{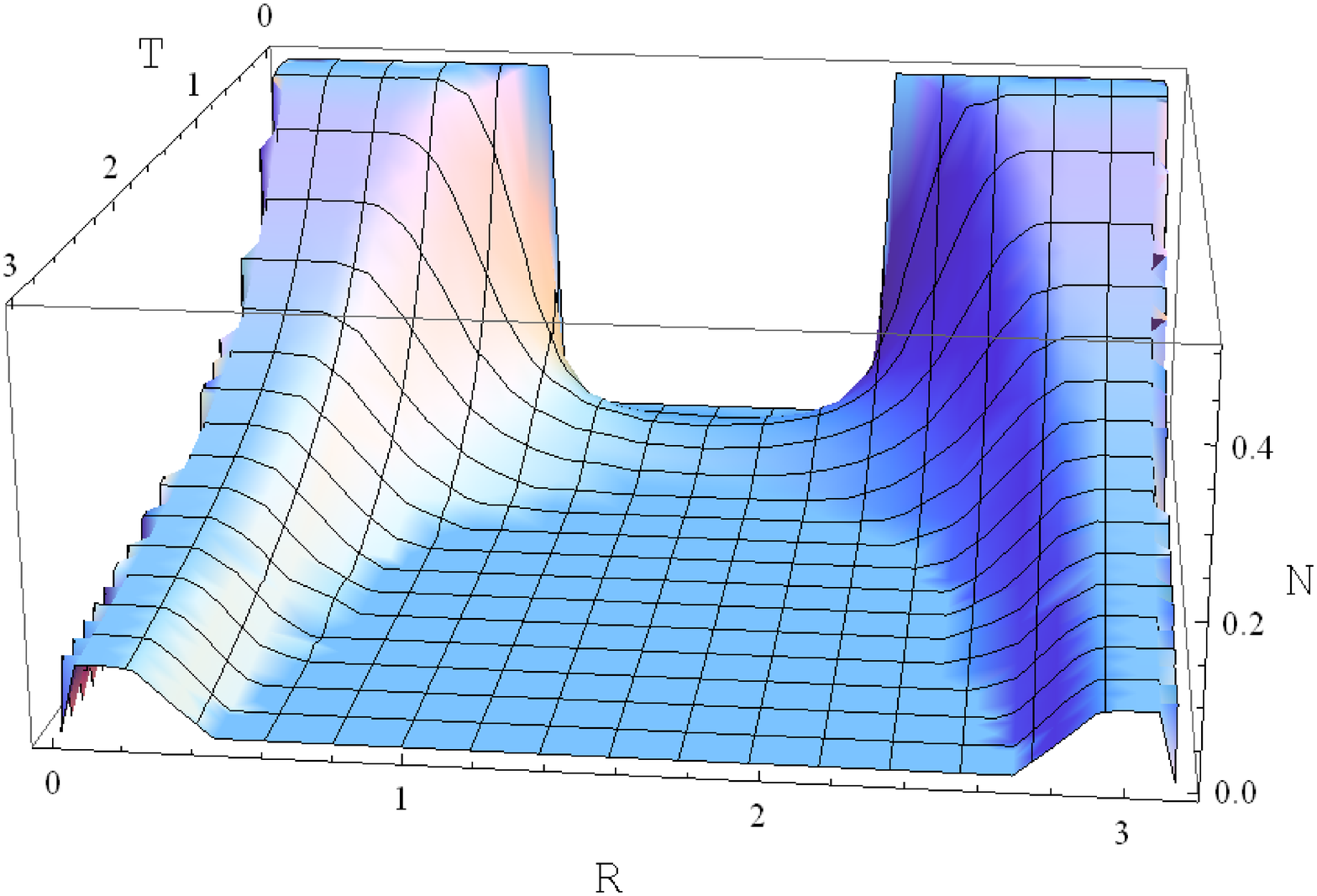}
\caption{Surface plot of the negativity for the two-spin
ferromagnetic XY system in the case $J(R)=1/\sin ^{2}R$ as a
function of temperature $T$ and interaction parameter $R$ at fixed
$B=1$.}}
\end{figure}
\begin{figure} \label{fig7}
\centering{
\includegraphics[width=3.0in,height=2.5in]{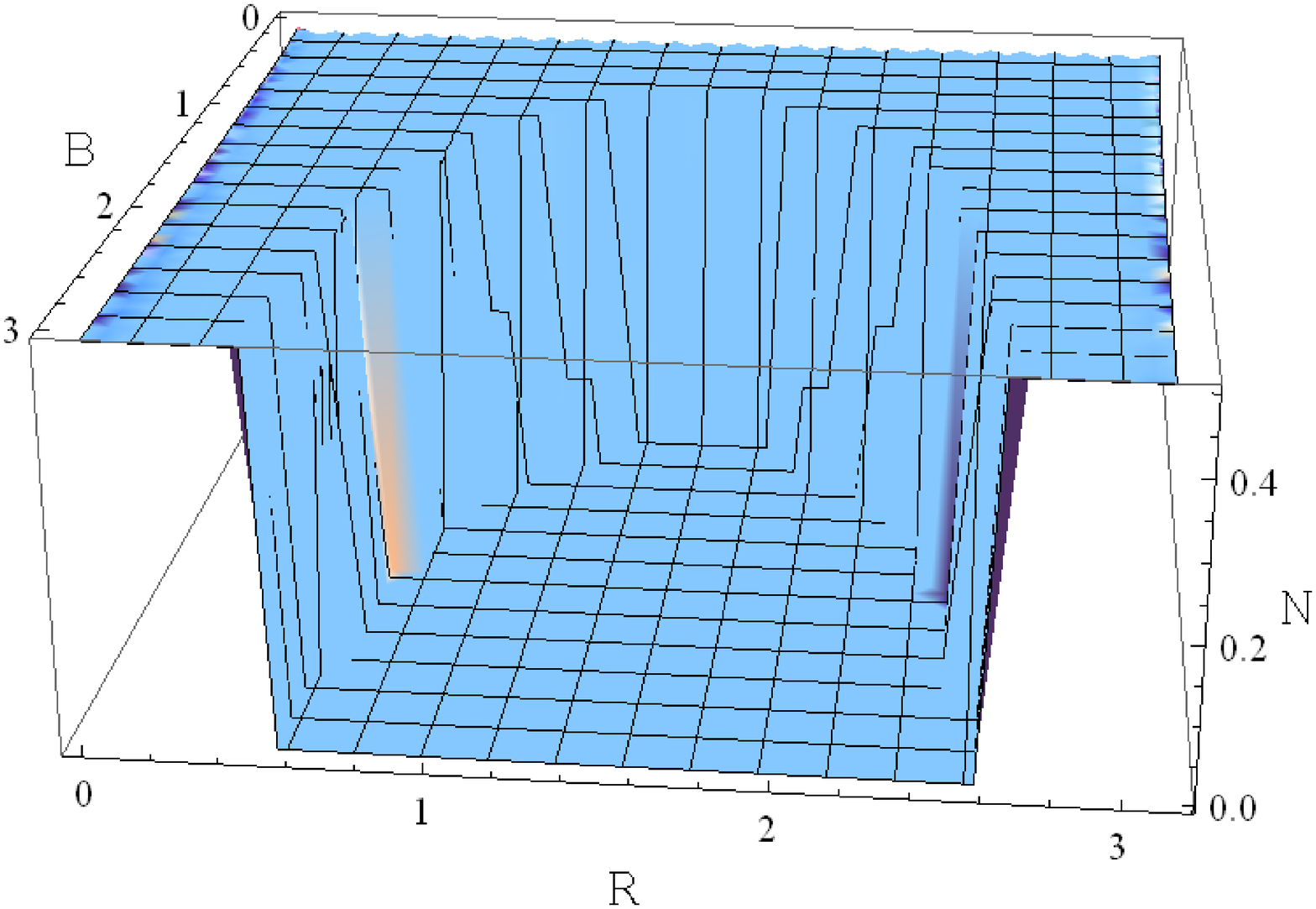}
\caption{Surface plot of the negativity for the two-spin
ferromagnetic XY system in the case $J(R)=1/\sin ^{2}R$ as a
function of magnetic field $B$ and interaction parameter $R$ at
fixed $T=0.001$.}}
\end{figure}
\begin{figure}
\centering{
\includegraphics[width=3.0in,height=2.5in]{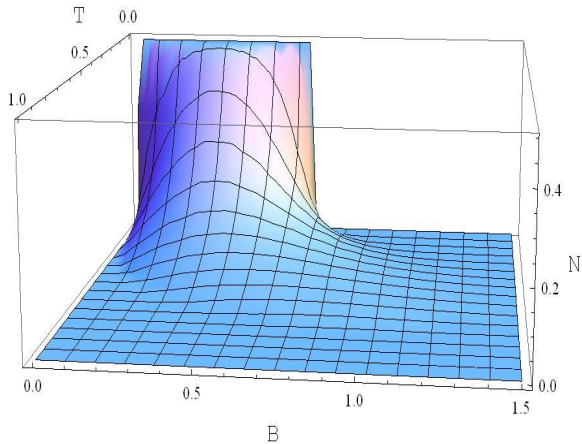}
\caption{Surface plot of the negativity for the two-spin
ferromagnetic XY system in the case $J(R)=1/\sin ^{2}R$ as a
function of temperature $T$ and magnetic field $B$ at fixed
$R=\pi/2$.}}
\end{figure}

In order to further demonstrate and unify the dependence of the
entanglement on $R$ ($0\leq R\leq \pi $) and $T$, the negativity
surface is plotted as a function of $R$ and $T$ for a fixed value
$B=1$ in Fig.\,(6). This surface plot shows the temperature
dependence of the interesting valley behavior of the negativity,
with side plateaus, as a function of $R$. It is clear that at this
fixed value of $B=1$, the negativity valley (symmetric in $R$ about
$R=\pi/2$) widens with increasing temperature, with the side
plateaus diminishing and dropping from their low $T$ maximum of
$N=0.5$. The $T$-dependence of the two symmetric critical $R_c$
values is clearly visible in this surface plot. However, whilst the
valley profile of the negativity with $R$ persists all the way down
in $T$ to zero temperature, it is clear that at the lowest
temperatures the valley floor exhibits a ridge for which the
negativity is non-zero for all values of $R$ including the mid-point
$R=\pi/2$. At these low temperatures critical behavior with $N$
dropping suddenly to zero thus disappears, although the negativity
still exhibits a critical change from the plateau at $N=0.5$ to the
small finite value at the valley floor. The behavior of this low-$T$
ridge in negativity as the temperature actually approaches zero is
dependent upon the fixed value of $B$ chosen, and will be
illustrated later in Fig.\,(8). Similarly to Fig.\,(5) singularity
behavior in negativity appears in the limit $R\rightarrow 0$ and
$R\rightarrow \pi$ in Fig.\,(6).

Similarly, in order to further demonstrate and unify the dependence
of the entanglement on $R$ ($0<R<\pi$) and $B$, the negativity
surface is plotted as a function of $R$ and $B$ for a fixed value
$T=0.001$ in Fig.\,(7). This surface plot shows the field dependence
of the interesting valley behavior of the negativity, with side
plateaus, as a function of $R$. It is clear that at this fixed value
of $T=0.001$, the negativity valley (symmetric in $R$ about
$R=\pi/2$) widens with increasing temperature, with the side
plateaus essentially remaining at their low $B$ maximum of $N=0.5$.
The $B$-dependence of the two symmetric critical $R_c$ values is
clearly visible in this surface plot and it is also clear that these
merge at a specific $B$ point and below this value of $B$ the
entanglement is at its maximum for all values of $R$, with no
critical behavior.

The final figure for the SCM interaction $J(R)=1/\sin ^{2}R$ is
Fig.\,(8), which shows the behavior with $B$ and $T$ of the
negativity surface at a fixed value of $R=\pi /2$, which is the
mid-point of the $R$ range---about which the valley behavior in $R$
in Fig.\,(6) and Fig.\,(7) is symmetric. It can be seen that
non-zero negativity appears at low $T$ for this fixed $R=\pi /2$,
consistent with the ridge seen in Fig.\,(6). The behavior of the
ridge (or valley floor) mid-point as a function of $B$ and $T$ is
captured in Fig.\,(8). It can be seen from this that at very low and
high magnetic field values, the negativity (and thus the ridge)
disappears as $T \rightarrow 0$ or may not even appear at all,
whereas for intermediate $B \sim 1/2$ the negativity (ridge) appears
and grows as $T \rightarrow 0$.

As a result, in this subsection, we have investigated the behavior
of the entanglement of a two-spin XY system in the case of the
long-range SCM interaction $J(R)=1/\sin ^{2}R$. We have discussed
the dependence of the negativity on interaction parameter $R$,
magnetic field $B$ and temperature $T$. We have demonstrated the
symmetric behavior about $R=\pi/2$ (with singular behavior as $R
\rightarrow 0$ and $R \rightarrow \pi$). The critical plateau and
valley behavior is seen in significant parameter regimes along with
the associated entanglement sudden death.

\subsection{SCM Type III: $J(R)=J_{0}/\sinh ^{2}R$ }

Finally we consider hyperbolic version of long-range interaction in
SCM model, which is defined with exchange interaction
$J(R)=J_{0}/\sinh ^{2}R$
\cite{Calogero,Sutherland,Moser,Gibbons,Wojciechowski,Haldane1,Shastry1}.
Here we set $J_{0}=1$ and $\theta=R$ for simplicity. The eigenvalues
for this type of SCM interaction are given by
\begin{eqnarray}
&&\pm 3B/2, \quad W_{\pm }=\frac{1}{2}(\pm
B+\frac{\sqrt{2}}{\sqrt{\sinh [R]^{4}}}),\nonumber\\ &&
Q_{\pm}=\frac{1}{2}(\pm B-\frac{\sqrt{2}}{\sqrt{\sinh [R]^{4}}}) \ .
\end{eqnarray}
Using the eigenvalues in Eq.\,(14) and the eigenvectors in Eq.\,(2)
(with Eq.\,(3)) the density matrix $\rho$ can be constructed.  For
the SCM interaction type of $J(R)=1/\sinh ^{2}R$, the partial
transpose of $\rho$ shown in Eq.\,(4) has matrix elements given in
Eq.\,(15) as follows:
\begin{subequations}
\begin{eqnarray}
a_{11}=\frac{1}{Z}e^{\frac{3B}{2T}}
\\
a_{12}=a_{21}=\frac{\eta}{2Z}(1-e^{\frac{B}{T}})
\\
a_{22}=a_{33}=\frac{\xi}{2Z}(1+e^{\frac{B}{T}})
\\
a_{44}=a_{55}=\frac{\eta}{2Z}(1+e^{\frac{B}{T}})
\\
a_{56}=a_{65}=\frac{\xi}{2Z}(1-e^{\frac{B}{T}})
\\
a_{66}=\frac{1}{Z}e^{-\frac{3B}{2T}}
\end{eqnarray}
\end{subequations}
where $\eta =\exp (-(B-\sqrt{2}$csch$[R]^{4}\sqrt{\sinh
[R]^{4}})/2T)$ , $\xi =\exp (-(B+\sqrt{2}/\sqrt{\sinh
[R]^{4}})/2T)$. The partition function for this case is simply given as
\begin{equation}
Z=2\cosh [\frac{3B}{2T}]+(1+e^{B/T})(\xi +\eta ) \ .
\end{equation}
Similarly, the negativity follows from summation of the negative
eigenvalues of the partial transpose density matrix Eq.\,(4) with
elements as in Eq.\,(15).

The negativity for this $J(R)=1/\sinh ^{2}R$ case is plotted versus
$R$ ($0< R\leq \pi $) for different temperature $T$ values at fixed
magnetic field value $B$ in Fig.\,(9a), and for different magnetic
field values at fixed temperature $T$ in Fig.\,(9b). It can be seen
from Fig.\,(9a) that, at a fixed value of $B=1$, the negativity
attains its maximum value in a plateau  for low temperature,
dropping suddenly to zero with increasing $R$ at a critical $R_c$
value, but it also drops to zero in the limit $R\rightarrow 0$. In
this $R \rightarrow 0$ limit, the negativity has a singularity. The
value of negativity on the plateau and the plateau width in $R$ both
vary with $T$, with the lowest value of $T=0.001$ exhibiting
saturation at $N=0.5$. Fig.\,(9b) also shows plateau behavior of
negativity at small $R$ with a sudden drop as $R$ exceeds a
threshold $R_c$. Similarly to Fig.\,(9a), in the limit $R\rightarrow
0$, a singularity appears in negativity. Negativity also drops to
zero in the this limit. However, for the fixed value $T=1$ chosen
none of the plateau values for $N$ attain the maximum, with the
value at the plateau increasing with increasing magnetic field $B$.
The plateau width in $R$ is essentially independent of $B$ for the
range considered.
\begin{figure}
\centering{
\includegraphics[width=3.5in,height=2.5in]{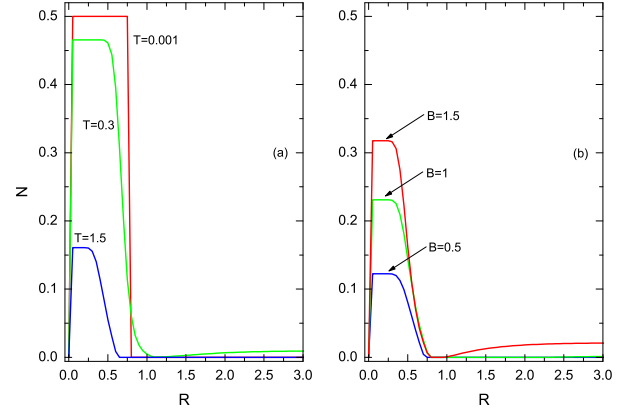}
\caption{Negativity as a function of $R$ for the two-spin
ferromagnetic XY model in the case $J(R)=1/\sinh ^{2}R$ (a) for
selected temperatures at fixed $B=1$, (b) for selected magnetic
fields at fixed $T=1$.}}
\end{figure}

In order to further demonstrate and unify the dependence of
entanglement on $R$ and $T$, the negativity surface is plotted as a
function $R$ $(-\pi \leq R\leq \pi $) and $T$ for a fixed value of
$B=1.5$ in Fig.\,(10). This surface plot further illustrates the
interesting plateau and critical behavior of the negativity. It can
be seen from Fig.\,(10) that for increasing $|R|$ the negativity
suddenly drops to zero at a critical $R_c$ value, but it also drops
to zero in the limit $R\rightarrow 0$. In this $R \rightarrow 0$
limit, the negativity has a singularity since $R$ values are chosen
between interval $-\pi \leq R\leq \pi$. This Fig.\,(10) clearly
demonstrates that the (symmetric positive and negative) critical
$R_c$ values are essentially independent of temperature $T$,
although the actual value of the negativity on the entanglement
plateau does decrease with increasing $T$.
\begin{figure}
\centering{
\includegraphics[width=3.0in,height=2.5in]{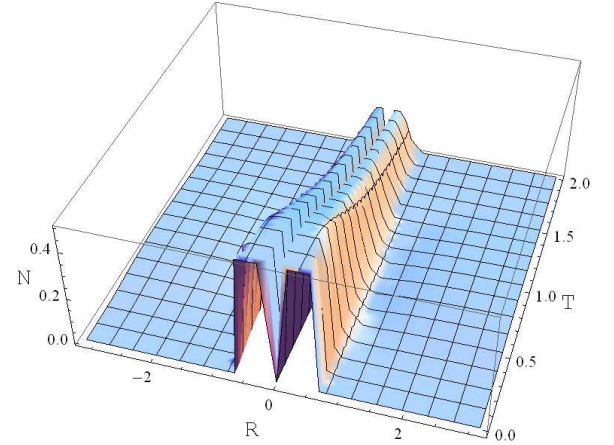}
\caption{Surface plot of the negativity for the two-spin
ferromagnetic XY system in the case $J(R)=1/\sinh ^{2}R$ as a
function temperature $T$ and interaction parameter $R$ at fixed
$B=1.5$.}}
\end{figure}
\begin{figure}
\centering{
\includegraphics[width=3.0in,height=2.5in]{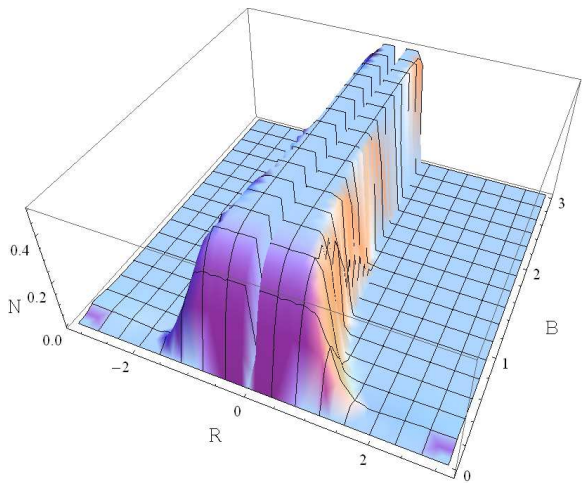}
\caption{Surface plot of the negativity for the two-spin
ferromagnetic XY system in the case $J(R)=1/\sinh ^{2}R$  as a
function magnetic field $B$ and interaction parameter $R$ at fixed
$T=0.1$.}}
\end{figure}

Finally, to further demonstrate and unify the dependence of
entanglement on $R$ and $B$, the negativity is plotted as a function
$R$ $(-\pi \leq R\leq \pi $) and $B$ for a fixed value $T=0.1$ in
Fig.\,(11). Again plateau and critical behavior is apparent. It can
be seen from Fig.\,(11) that the constant negativity on a plateau
suddenly drops to zero at a critical $R_c$ value. Here the
(symmetric positive and negative) $R_c$ show dependence on $B$,
whereas---aside from the small $B$ regime---the value of the
negativity on the plateau is essentially saturated at $N=0.5$.
Similarly to Fig.\,(10), in the limit $R \rightarrow 0$, a
singularity appears in the negativity since $R$ values are chosen
between interval $-\pi \leq R\leq \pi$. In the small $B$ regime the
critical behavior disappears and the negativity decreases smoothly
as the symmetric plateaus disappear for decreasing $B$ or increasing
$R$.

To summarize, in this final subsection,we have investigated the
behavior of the entanglement of a two-spin XY system in the case of
the long-range SCM interaction $J(R)=1/\sinh ^{2}R$. We have also
discussed the dependence of the negativity on interaction parameter
$R$, magnetic field $B$ and temperature $T$. We have demonstrated
the symmetric behavior about $R=0$ (with singular behavior as $R
\rightarrow 0$). Critical plateau behavior is seen over a wide
parameter regime along with the associated entanglement sudden
death.

\section{\label{sec:conclusions}conclusions}
\label{sec:con}

In this study, using the concept of negativity we investigate
entanglement in (1/2,1) mixed-spin XY models for the long-range
interaction with an inverse-square and its trigonometric and
hyperbolic variants given in the SCM model. We have also explored in
detail the temperature and magnetic field dependence of the thermal
entanglement in the (1/2,1) mixed-spin XY system for different types
of interactions. Our numerical results show that in the presence of
the SCM type interactions characterized by an interaction parameter
$R$, thermal entanglement between spin qubits has a rich behavior
dependent upon $R$, the temperature $T$ and the applied magnetic
field strength $B$. Indeed, we have found that there are
entanglement plateaus which saturate at the maximum negativity
$N=0.5$ for significant regions of the parameter ranges of $R$, $T$
and $B$. Aside from specific examples, in general the entanglement
plateaus are characterized by a critical distance $R_c$ and moving
off a plateau through $R_c$ demonstrates a critical change and
entanglement sudden death. Clearly the effect of SCM type
interactions on the resource of entanglement provides a rich source
of behavior, with maximum entanglement existing over significant
parameter regimes and critical switching.

The plateau behaviors in entanglement formation of two-spin given by
Eq.\,(1) for all types of SCM interactions are very interesting. The
physical mechanism of entanglement plateaus probably may be
long-range interactions. Indeed, while the plateau behavior in
entanglement does not appear for short-range interactions, here, we
can see that all long-range interactions with an inverse-square and
its trigonometric and hyperbolic variants lead to different plateau
behavior in the entanglement. Therefore, we can suggest that the
plateau behavior of entanglement in these model emerge from the
long-range character of the interactions. In all case for the
long-range SCM interactions, entanglement between spins still
subsists throughout plateau up to critical distance although spins
are spatially separated from each other unlike the system with
short-range interactions \cite{Venuti3,Osborne,Osterloh,Korepin}. We
believe that these interesting results can be valuable for
researcher in this area of physics.

\acknowledgements

This work was partially supported by YOK (The Council of Higher
Education), the University of Leeds and TUBITAK (The Scientific and
Technological Research Council of Turkey) under research project
(No.109T681). The author (EA) gratefully acknowledges YOK, the
University of Leeds, Istanbul University and TUBITAK. EA also
gratefully acknowledges to Viv Kendon for invitation as a visitor to
the Quantum Information Group (University of Leeds) where the first
stages of this work were completed during the visit.



\begin{thebibliography}{99}
\bibitem{sdcomment} Entanglement sudden death was first noted for evolution as a function of time. The usual use of this term applies to scenarios where the entanglement vanishes with a finite slope, rather than exhibiting an exponential or Gaussian decay, as a function of the relevant parameter.
\end{thebibliography}

\end{document}